%% file: paper2.tex
\documentclass[twocolumn,showpacs,aps,prl,superscriptaddress,nofootinbib]{revtex4}

\usepackage{graphicx}

\usepackage{dcolumn}

\usepackage{epsfig}

\input pubboard/babarsym.tex

\def\bldmth{\boldmath}  
\def\beq{\begin{equation}}
\def\eeq{\end{equation}}

\def\beqa{\begin{eqnarray}}
\def\eeqa{\end{eqnarray}}

\def\DKgood{{\ensuremath{DK_{\rm sig}}}}
\def\DKbad {{\ensuremath{DK_{\rm bgd}}}}

\def\Dpigood{{\ensuremath{D\pi_D}}}
\def\Dpibad {{\ensuremath{D\pi_{\not D}}}}
\def\Dpix   {{\ensuremath{D\pi X}}}
\def\DpiX   {\Dpix}
\def\DKX    {{\ensuremath{DKX}}}
\def\BBgood {{\ensuremath{BBC_D}}}
\def\BBbad {{\ensuremath{BBC_{\not D}}}}
\def\qqgood {{\ensuremath{qq_D}}}
\def\qqbad {{\ensuremath{qq_{\not D}}}}

\def\DE{\DeltaE}

\def\mes{\ensuremath{m_{ES}}}
\def\mD{\ensuremath{M_D}}
\def\mDst{\ensuremath{M_{D^*}}}
\def\NNqq{\ensuremath{q}}

\def\NNcomb{\ensuremath{d}}

\def\mppm{\ensuremath{s_{-}}}
\def\mppp{\ensuremath{s_{+}}}
\def\mpppm{\ensuremath{s_{\pm}}}
\def\mppmp{\ensuremath{s_{\mp}}}
\def\NR{\ensuremath{{\rm NR}}} 

\def\rhop{{\ensuremath{\rho^+}}}
\def\rhom{{\ensuremath{\rho^-}}}
\def\rhoz{{\ensuremath{\rho^0}}}

\def\pB{\ensuremath{\mathbf{p_B}}}

\def\P{\ensuremath{{\cal P}}}  
\def\E{\ensuremath{{\cal E}}}  
\def\DP{\ensuremath{{\cal D'}}}  
\def\DPP{\ensuremath{{\cal D}}}  
\def\Q{\ensuremath{{\cal Q}}}  
\def\C{\ensuremath{{\cal C}}}  
\def\B{{\cal B}}   
\def\fDz{\ensuremath{{\alpha}}}  

\def\A{{{ A}}}  

\def\eff{\ensuremath{\epsilon}}

\def\xz{{\ensuremath x_0}}

\def\zpm{{\ensuremath{z_{\pm}}}}

\def\Rp{{\ensuremath{\rho_+}}}
\def\Rm{{\ensuremath{\rho_-}}}
\def\Rpm{{\ensuremath{\rho_\pm}}}

\def\Tp{{\ensuremath{\theta_+}}}
\def\Tm{{\ensuremath{\theta_-}}}
\def\Tpm{{\ensuremath{\theta_\pm}}}

\def\btou{\ensuremath{b\to u\cbar s}}
\def\btoc{\ensuremath{b\to c\ubar s}}
\def\ppp{\ensuremath{\pi^+\pi^-\pi^0}}

\def\btodzk{\ensuremath{B^- \to \Dz K^-}}
 
\def\btodkgen{B\to D^{(*)0}K^{(*)}}
\def\bpmtodkpm{\ensuremath{B^\pm\to DK^\pm}}
 
\def\dtoppp{\ensuremath{D\to \ppp}}

\def\dztoppp{\ensuremath{\Dz\to \ppp}}

\def\dztokpp{\ensuremath{\Dz\to K^-\pi^+\piz}}
\def\Dppp{\ensuremath{D_{\pi^+\pi^-\piz}}}
\def\BDKppp{\ensuremath{B^- \to \Dppp K^-}}

\def\decaychain{\ensuremath{B^\pm \to \Dppp K^\pm}}

\def\bMtodpi{\ensuremath{B^- \to D^0 \pi^-}}

\def\BR#1{\ensuremath{{\cal B}(#1)}}

\newcommand{\BABARPubNumber}  {06/073}

\newcommand{\SLACPubNumber} {12410}

\def\figurebox#1#2#3{%
    \def\arg{#3}%
    \ifx\arg\empty
    {\hfill\vbox{\hsize#2\hrule\hbox to #2{\vrule\hfill\vbox to #1{\hsize#2\vfill}\vrule}\hrule}\hfill}%
    \else
    {\hfill\epsfbox{#3}\hfill}%
    \fi}

\begin{document}

\preprint{\BABARPubNumber} 
\preprint{SLAC-PUB-\SLACPubNumber} 

\begin{flushleft}
\babar-PUB-\BABARPubNumber\\
SLAC-PUB-\SLACPubNumber\\
\end{flushleft}

\title{\large \bf 
\boldmath
Measurement of \CP\ Violation Parameters with a Dalitz Plot Analysis of 
$\decaychain$
}

\input pubboard/authors_dec2006.tex


\date{\today}

\begin{abstract}
We report the results of a \CP\ violation analysis of the decay
$\decaychain$, where $\Dppp$ indicates a neutral $D$ meson detected in
the final state \ppp, excluding $\KS\piz$.
The analysis makes use of 324~million $\epem\to\BB$ events
recorded by the \babar\ experiment at the \pep2\ $\epem$ storage ring.
By analyzing the \ppp\ Dalitz plot distribution and the
$\decaychain$ branching fraction and decay rate asymmetry, we
calculate parameters related to the phase
$\gamma$ of the CKM unitarity triangle.  
We also measure the magnitudes and phases of the components of the 
$\Dz\to\ppp$ decay amplitude.
\end{abstract}

\maketitle
\vskip .3 cm


An important component 
of the program to study \CP~violation is the measurement of
the angle $\gamma = \arg{\left(- V^{}_{ud} V_{ub}^\ast/ V^{}_{cd}
V_{cb}^\ast\right)}$ of the unitarity triangle related to the 
Cabibbo-Kobayashi-Maskawa quark mixing matrix~\cite{ref:km}.
The decays $\btodkgen$ 
can be used to measure $\gamma$ with
essentially no hadronic uncertainties, exploiting interference
between \btou\ and \btoc\ decay amplitudes~\cite{Gronau:1991dp}. 
In one of the measurement methods~\cite{Giri:2003ty},
$\gamma$ is extracted by analyzing the $D$-decay Dalitz plot
distribution in  $\bpmtodkpm$
with multi-body $D$ decays~\cite{ref:D}.
This method has only been used with the Cabibbo-favored decay $D\to
\KS \pi^+\pi^-$~\cite{Abe:2003cn,Aubert:2004kv},
and Cabibbo-suppressed decays are expected to be similarly
sensitive to $\gamma$~\cite{Grossman:2002aq}.
We present here the first \CP-violation study of
$\bpmtodkpm$ with a multibody, Cabibbo-suppressed $D$ decay,	
$\dtoppp$. 
%


The data used in this analysis were collected with the \babar\
detector at the \pep2\  \epem\ storage ring, running
on the $\Upsilon(4{\rm S})$ resonance.
Samples of simulated Monte Carlo (MC) events were analyzed with the same
reconstruction and analysis procedures. These samples 
include an $\epem\to\BB$ sample
about five times larger than the data; a continuum $\epem\to\qqbar$ sample,
where $q$ is a $u$, $d$, $s$, or $c$ quark, with 
luminosity equivalent to the data; and a signal sample about 
300 times larger than the data, with both phase space 
$D$ decays and decays generated according to the
amplitudes measured by CLEO~\cite{cleoppp}.
The \babar\ detector and the methods used for particle
reconstruction and identification are described in 
Ref.~\cite{ref:babar}.

%
The reader is referred to
Ref.~\cite{Aubert:2005hi} for details of the event selection criteria.
Briefly, we use event-shape variables to suppress the continuum
background, and identify kaon and pion candidates using specific
ionization and Cherenkov radiation. 
The invariant mass of $D$
candidates must satisfy $1830 < \mD < 1895$~\mevcc.  We require
$5272 < \mes < 5300$~\mevcc, where $\mes \equiv \sqrt{E_{\rm CM}^2/4
- |\pB|^2}$, $E_{\rm CM}$ is the total \epem\ center-of-mass (CM) energy, and
\pB\ is the $B$ candidate CM momentum.  Events must satisfy 
$-70 < \DE < 60$~\mev, where
$\DE = E_B -
E_{\rm CM}/2$ and $E_B$ is the $B$ candidate CM energy.
We exclude the decay mode $D\to\KS\piz$,
which is a previously studied \CP\-eigenstate 
not related to the method of Ref.~\cite{Giri:2003ty}, 
by rejecting candidates with 
$489<M(\pi^+\pi^-)<508$~\mevcc or for which the
distance between the $\pi^+\pi^-$ vertex and the $B^-$ candidate decay
vertex is more than 1.5~cm.
We reject \decaychain\ candidates in which the $K^\pm \pi^\mp$ invariant mass 
satisfies $1840<M(K^\pm \pi^\mp)<1890$~\mevcc, to suppress 
$B^-\to D^0_{K^-\pi^+}\rho^-$ decays.
We require $\NNcomb > 0.25$, where 
\NNcomb~\cite{Aubert:2005hi}
is a neural net variable that
separates signal candidates (which peak toward $\NNcomb =
1$) from those with a misreconstructed $D$ (peaking toward $\NNcomb=0$).
In events with multiple candidates, we keep the candidate whose 
$\mes$ value is
closest to the nominal $B^\pm$ mass~\cite{ref:pdg06}.


For each \decaychain\ candidate, we compute the neural net
variable \NNqq~\cite{Aubert:2005hi}. The \NNqq\ distribution of
\BB\ events peaks toward $\NNqq=1$, while that of continuum peaks at
$\NNqq=0$.  
For $\nu \in\{ \NNqq, \NNcomb\}$, we define the variables
$\nu' \equiv \tanh^{-1}\left[(\nu-{1\over 2}(\nu_{max} + \nu_{min}))
        /{1\over 2}(\nu_{max} + \nu_{min})\right]$,
where $\NNqq_{max} = \NNcomb_{max} = 1$, $\NNqq_{min} = 0.1$, and $\NNcomb_{min} = 0.25$
are the allowed ranges for \NNqq\ and \NNcomb.  
The $\nu'$ variables can be conveniently fit with Gaussians,
as described later.


As in Ref.~\cite{Aubert:2005hi}, we identify in the MC samples
ten event types, one signal and nine different backgrounds. We list them 
here with the labels used to refer to them throughout the paper.
{\bldmth \DKgood}: \decaychain\ events that are correctly 
reconstructed; these are the only events considered to be signal.
{\bldmth \DKbad}: \decaychain\ events that are 
misreconstructed; namely, some of the particles used to form the final
state do not originate from the \decaychain\ decay.
{\bldmth \Dpigood} ({\bldmth \Dpibad}): $\bMtodpi$, $D^0\to\ppp$ decays, 
where the decay $D^0\to \ppp$ is correctly reconstructed (misreconstructed).
{\bldmth \DKX}: $B\to D^{(*)} K^{(*)-}$ events not containing the decay
$\dtoppp$.
{\bldmth \DpiX}: $B\to D^{(*)} \pi^-$ and $B\to D^{(*)}
\rho^-$ decays, excluding $\dtoppp$.
{\bldmth \BBgood} ({\bldmth \BBbad}): all other $\BB$ events with a
correctly reconstructed (misreconstructed) $D$ candidate.  
{\bldmth \qqgood} ({\bldmth \qqbad}){\bldmth}: 
continuum $\epem\to\qqbar$ events with a
correctly reconstructed (misreconstructed) $D$ candidate.  
%
%


The measurement of the \CP\ parameters proceeds in three steps, each
involving an unbinned maximum likelihood fit.  
In step~1, we
measure the complex Dalitz plot
amplitude $\fDz(\mppp,\mppm)$ for the decay \dztoppp,
where $\mpppm = m^2(\pi^\pm\pi^0)$ are the squared invariant masses of
the $\pi^\pm\pi^0$ pairs. 
In step~2, we 
extract the numbers of $B^+$ and $B^-$ signal events and 
background yields. 
We obtain the \CP\ parameters in step~3.

We parameterize $\fDz(\mppp,\mppm)$ using the
isobar model,
$\fDz(\mppp,\mppm) = [a_{\NR} e^{i\phi_\NR} + 
        \sum_r a_r e^{i\phi_r} A_r(\mppp, \mppm)]/N_\alpha$,
where the first term represents a nonresonant contribution, 
the sum is over all intermediate two-body resonances $r$, and
$N_\alpha$ is such that $\int d\mppp d\mppm |\fDz(\mppp,\mppm)|^2 = 1$.
The 
amplitude for the decay chain $\Dz\to rC$, $r\to{AB}$ 
is 
  $ A_r(\mppp, \mppm) = 
        F_r F_s \left(m_r^2-M_{AB}^2-im_r\Gamma_r(M_{AB})\right)^{-1}$,
where   $m_r$ is the peak mass of the resonance~\cite{ref:pdg06},
        $M_{AB}^2$ is the squared invariant mass of the $AB$ pair,
        $F_r$ is a spin-dependent form factor~\cite{Kopp:2000gv},
        and $\Gamma_r(M_{AB})$ is the mass-dependent width for the 
	resonance $r$~\cite{Kopp:2000gv}.
The spin factors $F_s$ are
$F_0=m_D^2$, 
$F_1 = M_{BC}^2-M_{AC}^2+(m_{D}^2-m_C^2)(m_A^2-m_B^2) {M_{AB}^{-2}}$,
and
$F_2 = \left(F_1^2 - {1 \atop 3}\mu^2_{CD}\,\mu^2_{AB}\right)  m_D^{-2}$,
where 
$\mu^2_{jk} \equiv M_{AB}^2 - 2m_j^2 - 2m_k^2 + \left(m_j^2 - m_k^2\right)^2 M_{jk}^{-2}$,
and $m_i$ is the mass of particle $i$~\cite{ref:pdg06}.


In step~1, we determine the parameters $a_{\NR}$, $a_r$, $\phi_{\NR}$, and
$\phi_r$ by fitting a large sample of $\Dz$ and $\Dzb$ mesons,
flavor-tagged through their production in the decay
$D^{*+}\to\Dz\pi^+$~\cite{ref:dppp}. To select this sample, we require
the CM momentum of the $D^*$ candidate to be greater than 2770~\mevc,
and $|\mDst - \mD - 145.4~\mevcc| < 0.6~\mevcc$, where \mDst\ is
the invariant mass of the $D^*$ candidate.
The signal and background yields are obtained from a fit to the
$\mD$ distribution, modeling the signal 
as a Gaussian and the background as an exponential. 
The signal Gaussian
peaks at $1863.7 \pm 0.4$~\mevcc and 
has a width of $17.4 \pm 0.8$~\mevcc. 

Of the $\Dz$ candidates in the signal region $1848 < \mD < 1880$~\mevcc,
we obtain from the fit $N_S = 44780 \pm 250$ signal 
and $N_B = 830 \pm 70$ background events.
To obtain the parameters of $\fDz(\mpppm,\mppmp)$, we fit these candidates 
with the probability distribution function (PDF) 
$\; N_S\, |\fDz(\mppp, \mppm)|^2 \eff(\mppp, \mppm)
	+ N_B\, |f_B(\mppp, \mppm)|^2$, 
where the background PDF $f_B(\mppp, \mppm)$ 
is a binned distribution obtained from
events in the sideband $1930<\mD<1990$~\mevcc,
and $\eff(\mppp,\mppm)$ is an efficiency function, parameterized as a
two-dimensional third-order polynomial determined from MC.
To within the MC-signal statistical uncertainty,
$\eff(\mppp, \mppm) = \eff(\mppm, \mppp)$.
The region $\mD < 1848$~\mevcc, which
contains $\Dz\to K^-\pi^+\pi^0$ events that are absent
from the signal region, is not used. 

\begin{figure}[!htbp]
  \begin{center}
    \begin{tabular}{cc} 
\includegraphics[width=0.235\textwidth]{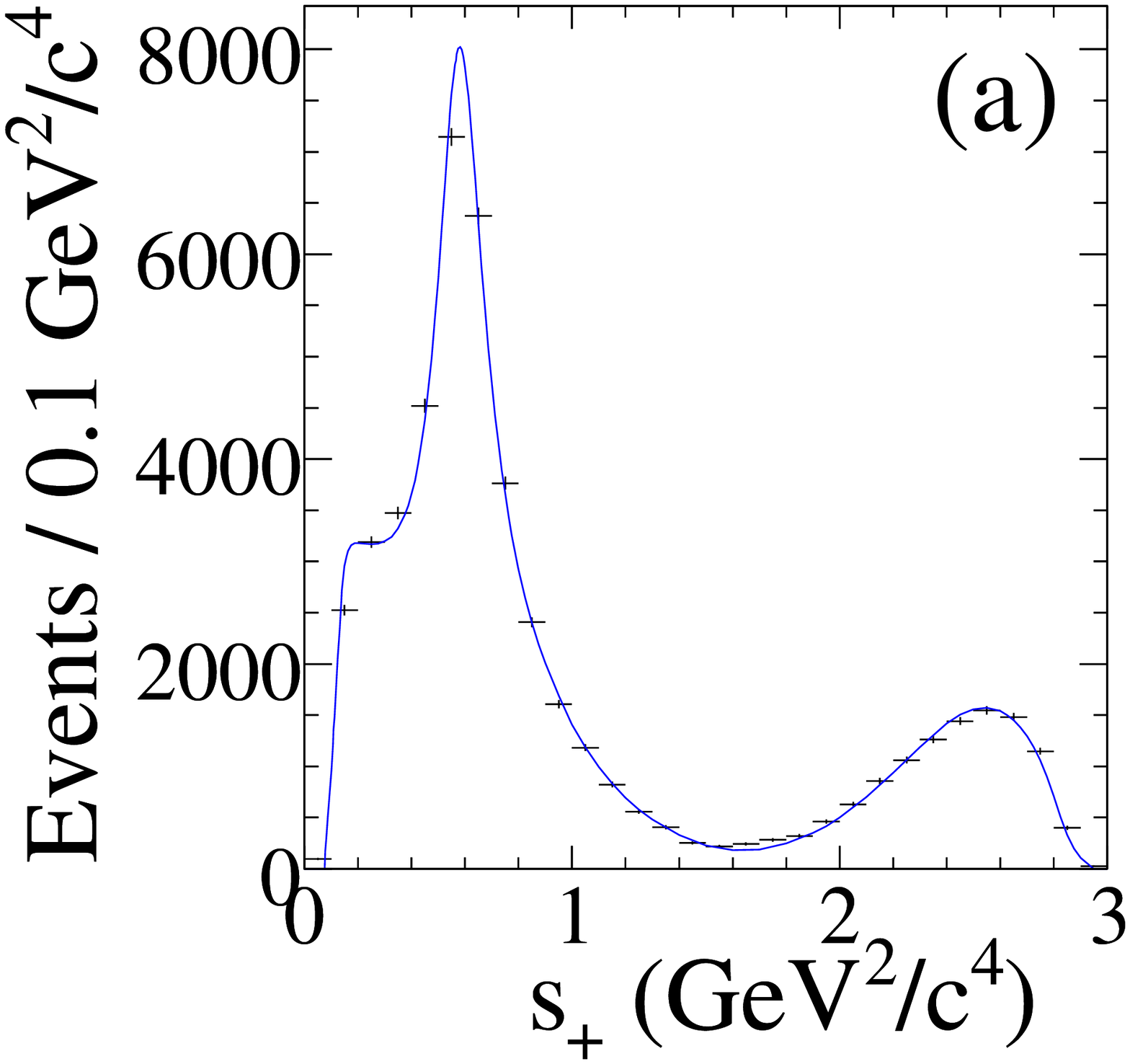} 
&
\includegraphics[width=0.235\textwidth]{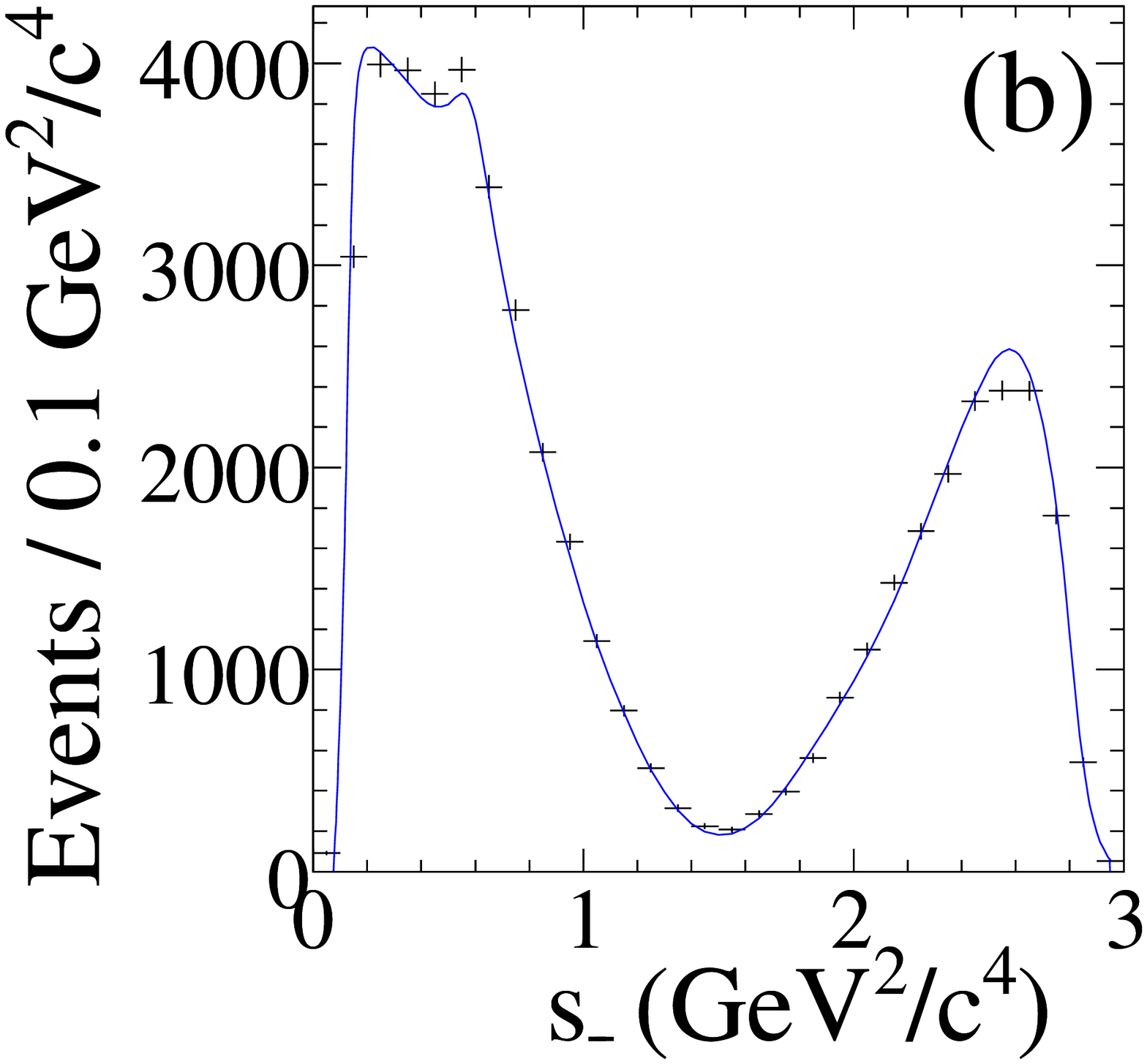} 
\\
\includegraphics[width=0.235\textwidth]{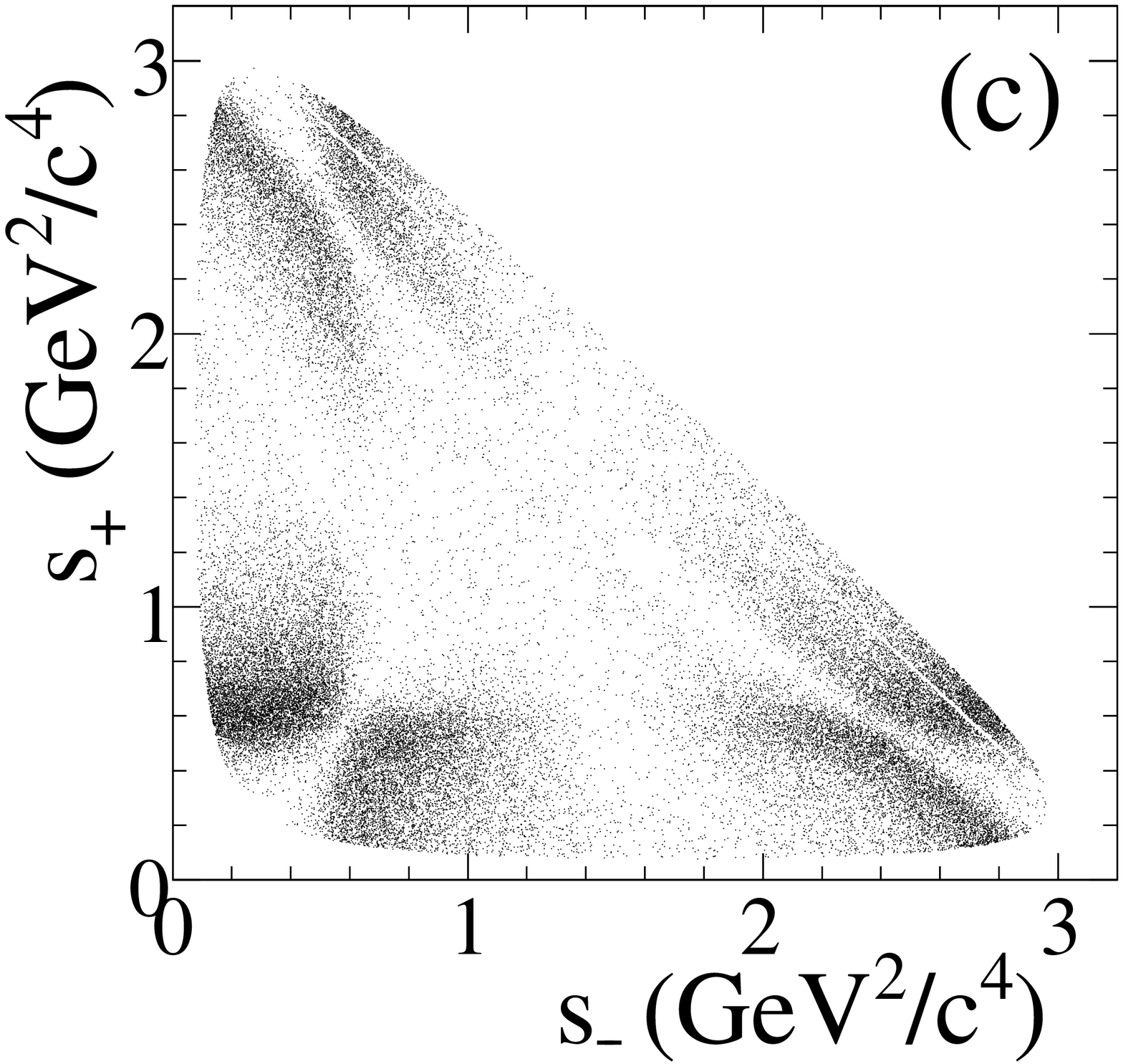}
&
\includegraphics[width=0.235\textwidth]{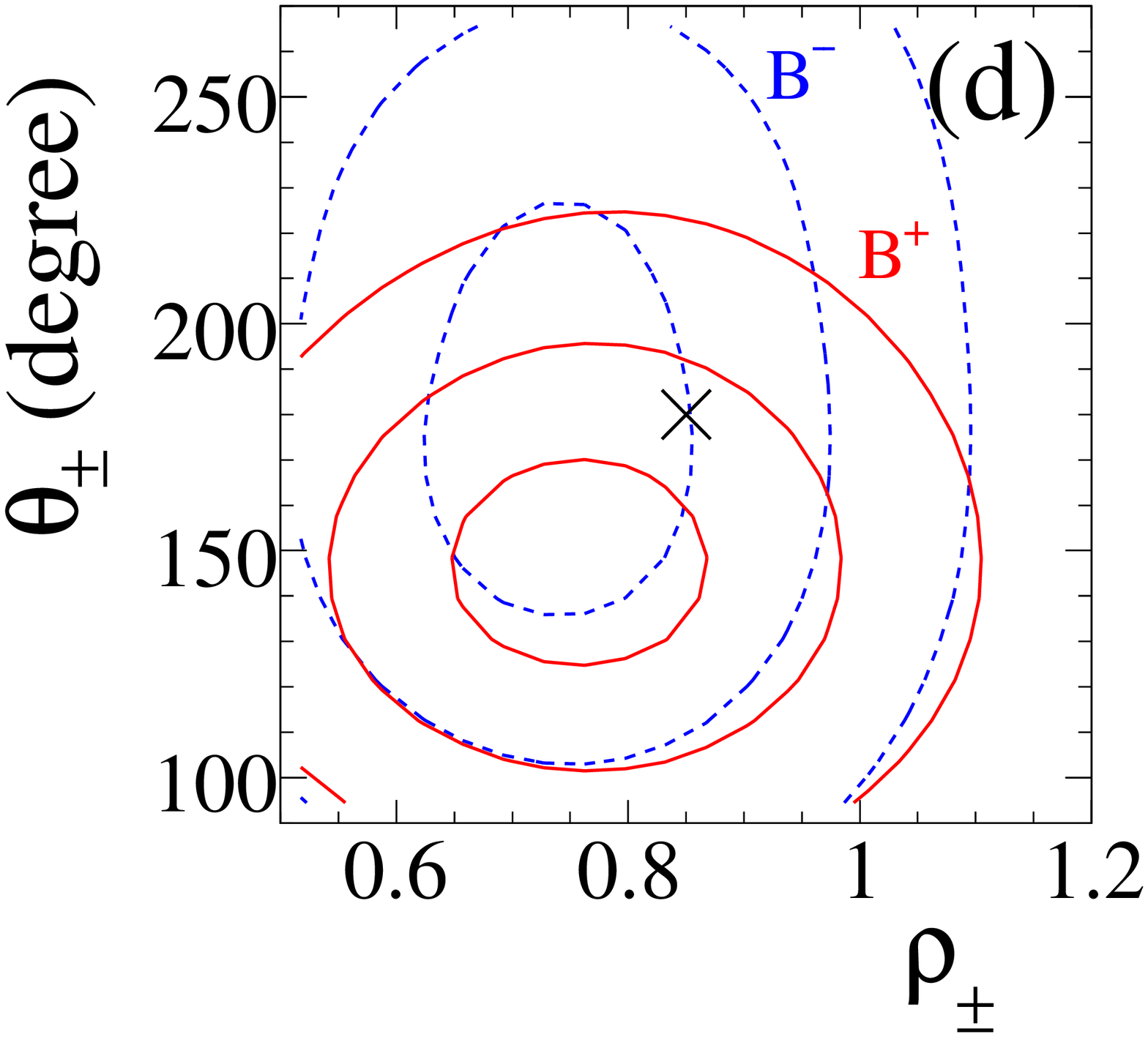} 
\\
\includegraphics[width=0.235\textwidth]{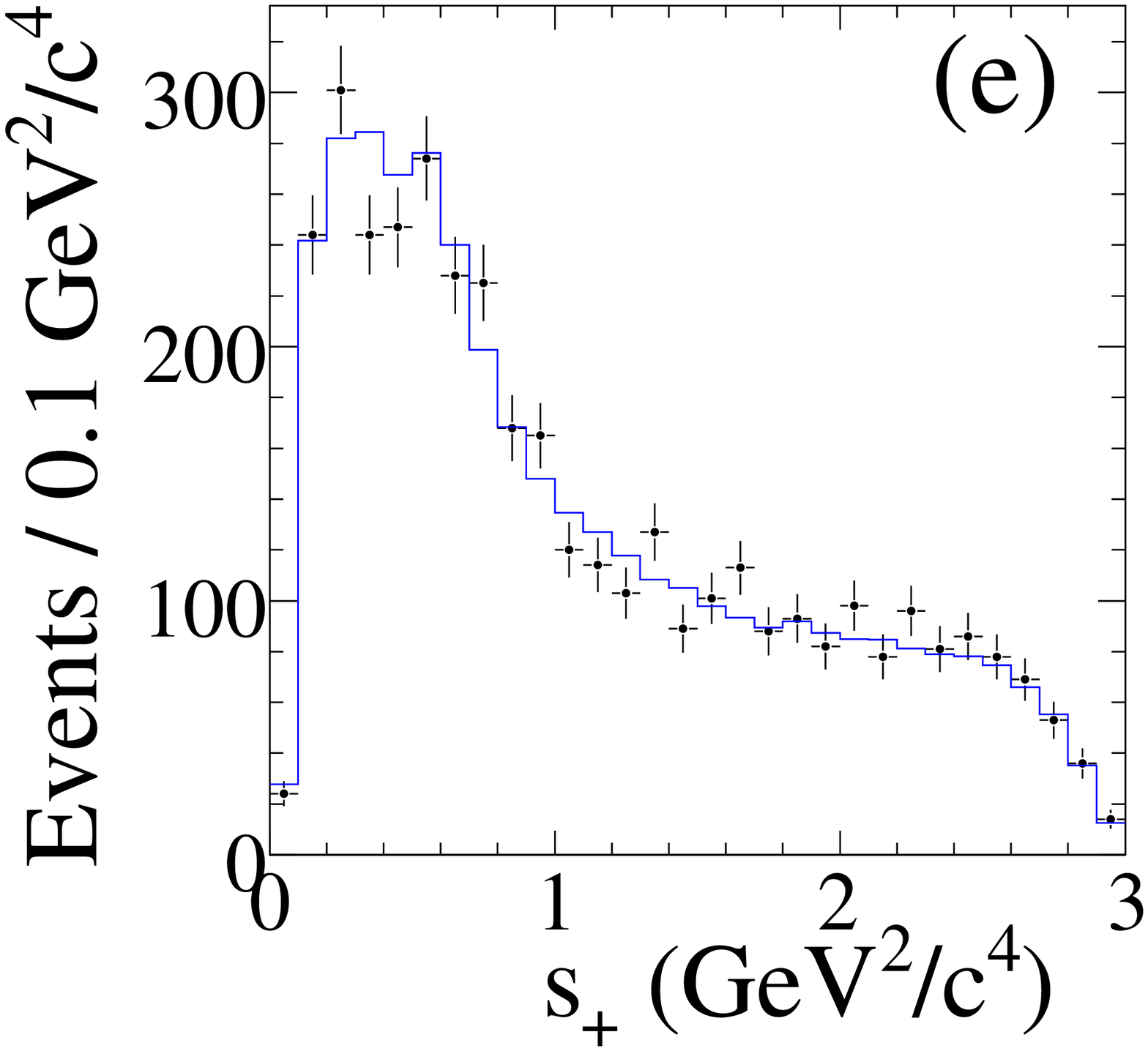} 
&
\includegraphics[width=0.235\textwidth]{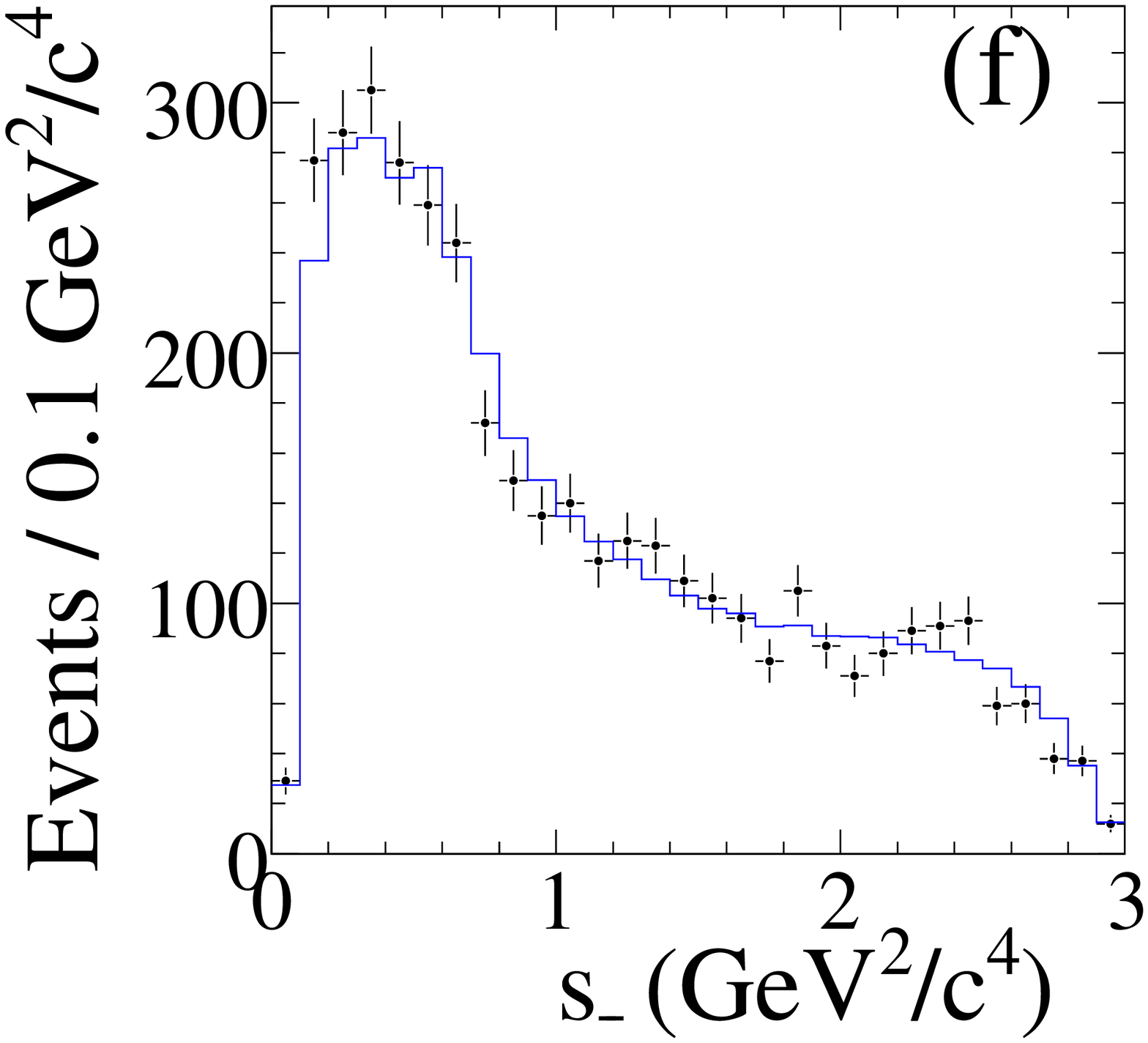} 
\\
\end{tabular}
\caption{ (Charge conjugation is implied for all plots.) 
(a,b) Projections of the $D^{*+}\to \Dz\pi^+$ data events 
and PDF onto the Dalitz plot variables \mppp\ and \mppm.
(c) The 2-dimensional $(\mppp, \mppm)$ 
	distribution of the $D^{*+}\to \Dz\pi^+$ data.
(d) One-, two-, and three-standard-deviation 
contours of $\calL$ as a function of \Tpm\ vs. 
\Rpm. The solid (dashed) curves correspond to $B^+$ ($B^-$) results. 
The no-interference point $(\Rpm=\xz, \Tpm=180^\circ)$ is marked
with an $\times$.
(e,f) Projection of the $\BDKppp$ candidate data onto 
\mppp\ and \mppm.}
\label{fig:results}
\end{center}
\end{figure}

Table~\ref{tab:DstarFit} summarizes the results of this fit, with
systematic errors
obtained by varying the masses and
widths of the $\rho(1700)$ and $\sigma$ resonances, 
setting $F_r=1$, and varying $\eff(\mppp, \mppm)$
to account for uncertainties in reconstruction and particle identification.
The Dalitz plot
distribution of the data is shown in
Fig.~\ref{fig:results}(a-c). 
The distribution is marked by 
three destructively interfering $\rho\pi$ amplitudes, 
suggesting an $I=0$-dominated final state~\cite{ref:zemach}.

\begin{table}[htbp]
\caption{\label{tab:DstarFit}Result of the fit to the $D^{*+}\to\Dz\pi^+$ 
	sample, 
        showing the amplitudes ratios $R_r \equiv a_r/a_{\rhop(770)}$, 
	phase differences $\Delta \phi_r \equiv \phi_r - \phi_{\rhop(770)}$,
	and fit fractions
        $f_r \equiv \int |a_r A_r(\mppp,\mppm)|^2 d\mppm d\mppp$. 
	The first (second) errors are statistical
        (systematic). We take the mass (width) of
        the $\sigma$ meson to be 400 (600)~\mevcc.}
   \centering
   \begin{tabular}{l|r|r|r}
     \hline\hline
     State & $R_r$ (\%) & $\Delta \phi_r$ ($^\circ$) & $f_r (\%)$\cr
     \hline
   $\rhop(770)$  &   100 &  0 &  67.8$\pm$0.0$\pm$0.6\cr
     $\rhoz(770)$  &   58.8$\pm$0.6$\pm$0.2 &  16.2$\pm$0.6$\pm$0.4 & 26.2$\pm$0.5$\pm$1.1\cr
     $\rhom(770)$  &   71.4$\pm$0.8$\pm$0.3 & $-$2.0$\pm$0.6$\pm$0.6  & 34.6$\pm$0.8$\pm$0.3\cr
     $\rhop(1450)$ &   21$\pm$6$\pm$13    &  $-$146$\pm$18$\pm$24    & 0.11$\pm$0.07$\pm$0.12\cr
     $\rhoz(1450)$ &   33$\pm$6$\pm$4    &  10$\pm$8$\pm$13       & 0.30$\pm$0.11$\pm$0.07\cr
     $\rhom(1450)$ &   82$\pm$5$\pm$4    &  16$\pm$3$\pm$3       & 1.79$\pm$0.22$\pm$0.12\cr
     $\rhop(1700)$ &   225$\pm$18$\pm$14    & $-$17$\pm$2$\pm$3       & 4.1$\pm$0.7$\pm$0.7\cr
     $\rhoz(1700)$ &   251$\pm$15$\pm$13    & $-$17$\pm$2$\pm$2       & 5.0$\pm$0.6$\pm$1.0\cr
     $\rhom(1700)$ &   200$\pm$11$\pm$7    & $-$50$\pm$3$\pm$3       & 3.2$\pm$0.4$\pm$0.6\cr
     $f_0(980)$ & 1.50$\pm$0.12$\pm$0.17 & $-$59$\pm$5$\pm$4 & 0.25$\pm$0.04$\pm$0.04\cr
     $f_0(1370)$ & 6.3$\pm$0.9$\pm$0.9 & 156$\pm$9$\pm$6 & 0.37$\pm$0.11$\pm$0.09\cr
     $f_0(1500)$ & 5.8$\pm$0.6$\pm$0.6 & 12$\pm$9$\pm$4 & 0.39$\pm$0.08$\pm$0.07\cr
     $f_0(1710)$ &11.2$\pm$1.4$\pm$1.7 & 51$\pm$8$\pm$7 & 0.31$\pm$0.07$\pm$0.08\cr
     $f_2(1270)$ & 104$\pm$3$\pm$21 & $-$171$\pm$3$\pm$4 & 1.32$\pm$0.08$\pm$0.10\cr
     $\sigma(400)$ & 6.9$\pm$0.6$\pm$1.2 & 8$\pm$4$\pm$8 & 0.82$\pm$0.10$\pm$0.10\cr
    Non-Res       &   57$\pm$7$\pm$8    &  $-$11$\pm$4$\pm$2      & 0.84$\pm$0.21$\pm$0.12\cr
     \hline\hline
   \end{tabular}
\end{table}


The fit for step $i\in\{2,3\}$ uses the PDF
\beq
\P^C_i  = 
  \sum_{t} {N_t\over2\eta}  {\left(1-C A_t\right)}\, 
         \P^{(C)}_{i,t}(\xi_i) 
	\div
		\int \P^{(C)}_{i,t}(\xi'_i)\, d^{n_i}\xi'_i ,
\eeq
where $\xi_i$ is the set of $n_i$ event variables 
$\xi_1 = \{\DE, \NNqq', \NNcomb'\}$, 
$\xi_2 = \{\DE, \NNqq', \mppm, \mppp\}$, 
$t$ corresponds to one of the ten event types
listed above, $N_{t}=N_t^+ + N_t^-$ is the number of events of type $t$, $A_t =
(N_t^- - N_t^+)/ N_t$ is their 
charge asymmetry, 
$C = \pm1$ is the electric charge of the $B$ candidate, and $\eta
\equiv \sum_t N_t$.
Using MC, we verify that the $\xi_i$ and $\xi_j$
($i\ne j$) distributions are uncorrelated for each event type. Therefore, 
the PDFs $\P^{(C)}_{i,t}$ are the products 
\begin{eqnarray} 
\kern-3mm \P_{2,t}(\DE, \NNqq', \NNcomb') &=&
          \E_t(\DE)\, \Q_t(\NNqq')\, \C_t(\NNcomb')\nonumber\\
\kern-3mm \P^C_{3,t}(\DE, \NNqq', \mppp, \mppm) &=&
          \E_t(\DE)\, \Q_t(\NNqq')\, \DP^C_t(\mppp, \mppm).
\label{eq:prodPDF}
\end{eqnarray}
%
The parameters of the Dalitz plot PDF 
$\DP^C_\DKgood(\mppp, \mppm)$ are obtained from the data
as described below.  Those of all other functions in
Eq.~(\ref{eq:prodPDF}) are obtained from the MC samples.
The functions $\E_t(\DE)$ are parameterized as 
the sum of a Gaussian and a second-order
polynomial. 
The PDFs $\Q_t(\NNqq')$ and $\C_t(\NNcomb')$ are the 
sum of a Gaussian and an asymmetric Gaussian.
The PDF parameters are different for each event type.
Assuming no \CP\ violation in the background,
we take
$\DP^{+}_t(\mppp, \mppm) = \DP^{-}_t(\mppm, \mppp)$ 
and $A_t=0$
for $t\ne\DKgood$.
The functions $\DP^C_\DpiX(\mppp, \mppm)$ and $\DP^C_\DKbad(\mppp, \mppm)$
are binned histograms obtained from the MC.
For other event types, $\DP^C_t(\mppp, \mppm) = \eff(\mppp, \mppm)
\DPP^C_t(\mppp, \mppm)$, where
the efficiency function $\eff(\mppp, \mppm)$ has different
parameters for well-reconstructed and misreconstructed $D$ candidates.

The signal Dalitz PDF accounts for interference between the \btou\ and
\btoc\ amplitudes $A_u$ and $A_c$:
\beq
\DPP^{\pm}_\DKgood(\mppp, \mppm) =   
\left| \fDz(\mppmp,\mpppm) + \zpm \fDz(\mpppm, \mppmp)  \right|^2,
\eeq
where $\zpm = |A_u/A_c| e^{i(\delta\pm\gamma)}$
and $\delta$ is a \CP-even phase.
 

In the step-2 fit, we extract the \decaychain\ signal yield and asymmetry,
as well as some background yields,
as described in Ref.~\cite{Aubert:2005hi}. 
From this fit we find
$N_\DKgood = 170 \pm 29$ signal events and a decay rate asymmetry 
$\A_\DKgood = -0.02 \pm 0.15$.
Errors are statistical only.


Only the complex parameters $\zpm$ are free in the step-3 fit. This fit
minimizes the function
\beq
 \calL = -\sum_{e=1}^{N_{\rm ev}} 
	\log \P_3^{C_e}(\xi_3^e) + {1 \over 2} \chi^2,
\label{eq:nll}
\eeq
where $N_{\rm ev}$ is the number of events in the data sample.
The term
$\chi^2 = \sum_{u,v=1}^2 X_u V_{uv}^{-1} X_v$
increases the sensitivity of the fit by using the results of 
the step-2 fit via
\beqa
X_1 &=& N_\DKgood - (n_- + n_+), 
        \nonumber\\
X_2 &=& A_\DKgood - (n_- - n_+) / (n_- + n_+),
\eeqa
where 
\beq
n_{\pm} = N^0 
        {\int\DP^{\pm}_\DKgood (\mppp, \mppm) d\mppp d\mppm
        \over   \int|\fDz(\mppmp, \mpppm)|^2 \eff(\mppp,\mppm)d\mppp d\mppm}
\label{eq:exp-yield}
\eeq
are the expected numbers of $B^\pm$ signal events. In Eq.~(\ref{eq:exp-yield}),
$N^0$ is the product of the number $N_{\Bp\Bm}$ 
of charged $\Bp\Bm$ pairs in the dataset, the branching
fractions $\BR{\btodzk}$~\cite{ref:pdg06} and $\BR{\dztoppp}$~\cite{ref:dppp},
and the total reconstruction efficiency $\epsilon = 11.4\%$.
The error matrix $V_{uv}$
is the sum of two components:
the step-2 fit error matrix $V_{uv}^{\rm stat}$, which is almost
diagonal (the correlation coefficient is 
$-2.8\%$), 
and the $N^0$ systematic error matrix $V_{uv}^{\rm syst}$. 
Here $V_{12}^{\rm syst} = V_{22}^{\rm syst} = 0$, and
$V_{11}^{\rm syst} = \sum_{c=1}^4 (N^0 \, \sigma_c^{\rm rel})^2$,
where $\sigma_c^{\rm rel}$ are the relative errors on the four
components
$N_{\Bp\Bm}$ (1.1\%), 
$\epsilon$ (3.3\%),
$\BR{\dtoppp}$ (3.8\%)~\cite{ref:dppp}, 
and $\BR{\btodzk}$ (5.9\%)~\cite{ref:pdg06}.


We parameterize \zpm\ with the polar coordinates
\beq
\rho_\pm \equiv |\zpm - \xz|,  \ \ \
   \Tpm \equiv \tan^{-1}\left({\Im [\zpm] \over \Re [\zpm] - \xz}\right),
\label{eq:polar-def}
\eeq
where $\xz$ is a coordinate transformation parameter,
\beq
\xz \equiv -\int \Re\left[\fDz(\mppp, \mppm) \,
                \fDz^*(\mppm, \mppp)\right] d\mppp d\mppm = 0.850. 
\eeq
This parameterization is optimal due to the polar symmetry of
$n_\pm =  N^0(1 + \Rpm^2 - \xz^2)$. Other parameterizations,
such as $(|A_u/A_c|, \g, \delta)$ or $(\Re[\zpm], \Im[\zpm])$, 
result in significant nonlinear correlations between the 
fit variables, which cannot be parameterized with an error matrix,
and bias the fit result.
The polar coordinates enable a significant improvement in 
sensitivity due to the $\chi^2$ term in Eq.~(\ref{eq:nll}),
and are determined from parameterized simulation to be 
unbiased.
The step-3 fit yields
\beqa
\kern-3mm  \Rm = 0.72 \pm 0.11 \pm 0.04, 
        & &\Tm = (173\pm 42 \pm 2)^\circ, \nonumber\\
\kern-3mm  \Rp = 0.75 \pm 0.11 \pm 0.04, 
        & &\Tp = (147 \pm 23 \pm 1)^\circ,
\label{eq:results-step3-stat}
\eeqa
where the first errors are statistical and the second are systematic,
due only to $V_{11}^{\rm syst}$.  
The largest correlation
coefficient is $c_{\Rm\Rp} = 14\%$, originating from $V_{11}^{\rm syst}$.  
All others are 1\% or less. 
Contours of constant $ \calL$ values are shown in 
Fig.~\ref{fig:results}(d). Projections of the data and the 
PDF onto \mppp\ and \mppm\ are shown in Fig.~\ref{fig:results}(e-f).

Additional systematic errors due to the analysis procedure are evaluated for 
the signal branching fraction, charge asymmetry, \Rpm, and \Tpm.
The uncertainty in the model used for $\fDz(\mppp,\mppm)$ 
is the largest source of
error on the \CP\ parameters: $\sigma_{\Rpm}^{\rm model} = 0.03$,
$\sigma_{\Tm}^{\rm model} = 14^\circ$,
$\sigma_{\Tp}^{\rm model} = 11^\circ$.
This error is evaluated by removing all but the
$\rho(770)$, $\rho(1450)$, $f_0(980)$, and nonresonant terms in 
$\fDz(\mppp, \mppm)$; adding an
$f_2'(1525)$, an $\omega$, and a nonresonant P-wave contribution; varying 
the meson ``radius'' parameter in $F_r$~\cite{Kopp:2000gv}; and propagating
the errors from Table~\ref{tab:DstarFit}.
Uncertainties due to the masses and widths of the $\rho(1700)$ and $\sigma$
resonances are small by comparison.
Other errors are due to uncertainties on
background yields that are fixed in the fits~\cite{Aubert:2005hi}, 
finite MC sample size,
a possible reconstruction efficiency charge asymmetry, and
uncertainties in the
background PDF shapes, evaluated by comparing MC and data in
signal-free sidebands of the variables \mD, \DE, and
$\mes$. 
We also evaluate errors due to possible charge asymmetries in \DKX\ and
\DKbad, uncertainties in particle identification and the efficiency
functions, the finite \mpppm\ measurement resolution, 
the background PDF $f_B$ in the $D^*$ sample, $D$-flavor mistagging in the 
$D^*$ sample, and correlations
between the $D$ flavor and the kaon charge in \qqgood\ events. 
These errors add in quadrature to
$\sigma^{\rm syst}_\Rpm = 0.05$, 
$\sigma^{\rm syst}_\Tm =19^\circ$, 
$\sigma^{\rm syst}_\Tp =13^\circ$,
and are combined with the systematic errors of 
Eqs.~(\ref{eq:results-step3-stat}).
%


The analysis procedure is validated in several ways. 
Conducting the analysis on the MC sample
yields results consistent with the generated values. 
We carry out the step-3 fit on a sample of 
$1800\pm 70$ $B^-\to \Dz_{\ppp}\pi^-$
events, obtaining the background Dalitz plot distribution from the
\DE\ sideband. The fit yields 
$\Rm = 0.815 \pm 0.034$, 
$\Tm = (186 \pm 7)^\circ$, 
$\Rp = 0.854 \pm 0.035$, 
$\Tp = (192 \pm 7)^\circ$, 
consistent with $\Rpm=\xz$, $\Tpm = 180^\circ$, which corresponds to $\zpm=0$.
We verify the signal efficiency by measuring the branching fraction
$\B(\bMtodpi)$ with \dztokpp\ and \dztoppp.
We compare the fit variable distributions of data and MC
events in signal-free sidebands. Good agreement is
found in all cases.


In summary, using a sample of $(324.0 \pm 3.6)\times 10^{6}$
$\epem\to\BB$ events, we observe $170 \pm 29$ \decaychain\ events.  We
calculate the branching fraction and decay rate asymmetry
\beqa
  \B(\decaychain) &=& (4.6 \pm 0.8 \pm 0.7)\times 10^{-6}, \nonumber\\
  \A(\decaychain) &=& -0.02 \pm 0.15 \pm 0.03,  
\eeqa
and the \CP-violation parameters
\beqa
\kern-4mm \Rm = 0.72 \pm 0.11 \pm 0.06, &   &\Tm = (173 \pm 42 \pm 19)^\circ, \nonumber\\
\kern-4mm \Rp = 0.75 \pm 0.11 \pm 0.06, &   &\Tp = (147 \pm 23 \pm 13)^\circ,
\eeqa
where the first errors are statistical and the second are systematic.
The parameters $\Rpm$, $\Tpm$ are defined in Eq.~(\ref{eq:polar-def}).
While the errors on $\Tpm$ are too large for a meaningful
determination of $\gamma$ with these results alone, our errors on
$\Rpm$ are small enough to make a non-negligible contribution to the overall
precision of $\gamma$ in a combination of all measurements related to
$\gamma$.
In addition, we measure the magnitudes and phases of the components of the 
amplitude of the decay $\Dz\to\ppp$ in the isobar model.


\input pubboard/acknow_PRL.tex

\end{document}

%% file: pubboard/authors_dec2006.tex
%
\author{B.~Aubert}
\author{M.~Bona}
\author{D.~Boutigny}
\author{Y.~Karyotakis}
\author{J.~P.~Lees}
\author{V.~Poireau}
\author{X.~Prudent}
\author{V.~Tisserand}
\author{A.~Zghiche}
\affiliation{Laboratoire de Physique des Particules, IN2P3/CNRS et Universit\'e de Savoie, F-74941 Annecy-Le-Vieux, France }
\author{E.~Grauges}
\affiliation{Universitat de Barcelona, Facultat de Fisica, Departament ECM, E-08028 Barcelona, Spain }
% ...
\collaboration{The \babar\ Collaboration}
\noaffiliation

%% file: pubboard/acknow_PRL.tex
We are grateful for the excellent luminosity and machine conditions
provided by our \pep2\ colleagues, 
and for the substantial dedicated effort from
the computing organizations that support \babar.
The collaborating institutions wish to thank 
SLAC for its support and kind hospitality. 
This work is supported by
DOE
and NSF (USA),
NSERC (Canada),
IHEP (China),
CEA and
CNRS-IN2P3
(France),
BMBF and DFG
(Germany),
INFN (Italy),
FOM (The Netherlands),
NFR (Norway),
MIST (Russia),
MEC (Spain), and
PPARC (United Kingdom). 
Individuals have received support from the
Marie Curie EIF (European Union) and
the A.~P.~Sloan Foundation.

%% file: paper2.bbl
\begin{thebibliography}{99}

\bibitem{ref:km} N.~Cabibbo, \jprl {\bf 10},
        531 (1963);
  M.~Ko\-ba\-yashi and T.~Maskawa, 
        Prog. Theoret. Phys.  {\bf 49}, 652 (1973).


\bibitem{Gronau:1991dp}
M.~Gronau and D.~Wyler, 
Phys.\ Lett.\ B {\bf 265},  172 (1991).


\bibitem{Giri:2003ty}
A.~Giri,  Y.~Grossman,  A.~Soffer and J.~Zupan, 
Phys.\ Rev.\ D {\bf 68},  054018 (2003);
A.~Bondar, Proceedings of BINP Special Analysis Meeting on Dalitz 
Analysis, 24-26 Sep. 2002, unpublished.



\bibitem{ref:D} We use the symbol $D$ to indicate any linear combination of
a $\Dz$ and a $\Dzb$ meson state. 

\bibitem{Abe:2003cn}
Belle Collaboration, A.~Poluektov {\it et al.},
  Phys.\ Rev.\ D {\bf 73}, 112009 (2006).



\bibitem{Aubert:2004kv}
\babar\ Collaboration, B.~Aubert {\it et al.},
  Phys.\ Rev.\ Lett.\  {\bf 95}, 121802 (2005).



\bibitem{Grossman:2002aq}
  Y.~Grossman, Z.~Ligeti and A.~Soffer,
  Phys.\ Rev.\ D {\bf 67}, 071301 (2003).


\bibitem{cleoppp}
See  ``Fit A'' in CLEO Collaboration, D.~Cronin-Hennessy {\it et al.},
  Phys.\ Rev.\ D {\bf 72}, 031102 (2005).


\bibitem{ref:babar}
\babar\ Collaboration, B.~Aubert {\it et al.}, 
Nucl.\ Instrum.\ Meth.\ A {\bf 479}, 1 (2002).


\bibitem{Aubert:2005hi}
\babar\ Collaboration, B.~Aubert {\it et al.},
  Phys.\ Rev.\ D {\bf 72}, 071102 (2005).


\bibitem{ref:pdg06}
Particle Data Group, Y.-M.~Yao {\it et al.},  
J.\ Phys.\ G {\bf 33}, 1 (2006).


\bibitem{Kopp:2000gv}
  See Eq.~(5) and Table~I in CLEO Collaboration, S.~Kopp {\it et al.},
  Phys.\ Rev.\ D {\bf 63}, 092001 (2001).


\bibitem{ref:dppp}
\babar\ Collaboration, B.~Aubert  {\it et al.}, 
  Phys.\ Rev.\ D {\bf 74}, 091102 (2006).

\bibitem{ref:zemach}
C.~Zemach, 
  Phys.\ Rev.\ {\bf 133}, B1201 (1964).


\end{thebibliography}
